# Singlet superconductivity in single-crystal NiBi$_3$ superconductor

G. J. Zhao[1,2], X. X. Gong[2], P. C. Xu[2], B. C. Li[1], Z. Y. Huang[1], X F. Jin[2], X. D. Zhu[3] and T. Y. Chen[1,2]*

[1]Institute for Quantum Science and Engineering and Department of physics of Southern University of Science and Technology, Shenzhen 518055, China

[2]Department of Physics, Arizona State University, Tempe, AZ 85287, USA

[3]State Key Laboratory of Surface Physics and Department of Physics, Fudan University, Shanghai 200433, China

[4]High Magnetic Laboratory, Chinese Academy of Sciences and University of Science and Technology of China, Hefei 230026, China

*correspondent author. E-mail: Tingyong.Chen@asu.edu

Abstract

Andreev reflection spectroscopy with unpolarized and highly spin-polarized currents has been utilized to study an intermetallic single-crystal superconductor NiBi$_3$. Magnetoresistance at zero bias voltage of point contacts shows the occurrence and suppression of Andreev reflection by unpolarized and polarized current, respectively. The gap value, its symmetry and temperature dependence have been determined using an unpolarized current. The spin state in the NiBi$_3$ sample is determined to be antiparallel using a highly spin-polarized current. The gap value $2\Delta/k_BT$, gap symmetry and its temperature dependence, combined with the antiparallel spin state show that the bulk NiBi$_3$ is a singlet *s*-wave superconductor.

**Introduction**

Superconductivity and ferromagnetism are two of the most amazing phenomena in solids. Below the Curie temperature of a ferromagnet (FM), exchange interaction aligns all atomic moments in the same direction. Spin of most or all conduction electrons in FMs points in the same direction and thus a current in a FM is spin-polarized, carrying spin angular momentum. For a superconductor (SC) below its transition temperature ($T_c$), however, two electrons form a Cooper pair, often with opposite spins. Such a supercurrent carried zero net spin angular momentum, in contrast to the spin-polarized current in a FM. Thus it seems that ferromagnetism and superconductivity are two antagonistic phenomena. Indeed, experimentally, it has been found that ferromagnetism is often detrimental to most SCs [1-5]. However, there is no fundamental obstruction for ferromagnetism and superconductivity to coexist. When two electrons form a bound state in a Cooper pair, the overall wavefunction must be antisymmetric, which requires the spin states to be singlet or triplet. In the singlet pairing, there is only one spin state, ⟨↑↓⟩ − ⟨↓↑⟩, whereas there are three spin states in a triplet pairing, ⟨↑↑⟩ ($S_z$ = 1), ⟨↑↓⟩ + ⟨↓↑⟩ ($S_z$ = 0), ⟨↓↓⟩ ($S_z$ = -1), with arrows denoting the spin states. Obviously, the two electrons in a singlet Cooper pair always have antiparallel spins while in a triplet SC, the two electrons can have parallel spins. Thus, ferromagnetism is compatible with *triplet* superconductivity where the supercurrent carries spin angular momentum, similar as that of a spin-polarized current in a FM.

Triplet superconductivity was predicted in the BCS theory about 60 years ago but a definitely verified triplet SC in solid state is rare. This is because almost all the macroscopic properties including Knight shift of singlet and triplet SCs are very similar and it is experimentally difficult to distinguish [6, 7]. Alternatively, coexistence of ferromagnetism and superconductivity has been utilized as an important technique to verify triplet superconductivity [8-17]. By measuring magnetic properties and Meissner effect, one can find the common temperature range of ferromagnetism and superconductivity. Such a SC is called ferromagnetic SC [9]. Using this technique, several SCs have been shown to have the coexistence of ferromagnetism and superconductivity including ErRh [10, 11], ErNi$_2$B$_2$C [12], UGe$_2$ [14], URhGe 15], NiBi$_3$ [17]. For magnetic SCs, the triplet nature often requires further verification because these materials often contain magnetic elements where sample uniformity and purity may also cause ferromagnetism in the samples [18-20]. Verification of triplet SCs, especially those that can carries spin angular momentum, is essential for understanding fundamental mechanism of unconventional superconductivity as well as important applications in quantum computing [21] and spintronics [22]. Since singlet and triplet SCs have similar macroscopic properties, an ideal experiment would be to directly measure the spin states of Cooper pairs in the SC.

Andreev reflection spectroscopy (ARS) using a highly spin-polarized or half-metallic current can directly assess the spin state of a Cooper pair. Spin polarization ($P$) describes the difference in the number of spin-up and spin-down electrons at the Fermi level, $P \equiv [N_\uparrow(E_F) - N_\downarrow(E_F)]/[N_\uparrow(E_F) + N_\downarrow(E_F)]$, with $E_F$ the Fermi level. A nonmagnetic metal has random spins thus $P$ = 0, while a half metal only has one spin band at the Fermi level, $P$ = 1. At a metal/SC interface, an injected electron from the metal must be accompanied by another electron with proper spin to go into the SC as a Cooper pair thus reflecting a hole into the metal. This process is called Andreev reflection (AR) [23]. For a normal metal with $P = 0$, each electron can find an appropriate partner to go into the SC as a Cooper pair. Consequently, the conductance within the gap is twice as that outside the gap. For a half metal, which carries a spin angular momentum, singlet Cooper pairs cannot be formed at the interface. Thus the conservation of spin angular moment prevents the injection of a half metallic current into a singlet SC. As a result, the conductance within the gap is zero. For ferromagnetic materials with $0 < P < 1$, the conductance is a linear combination of the two cases of $P = 0$ and $P = 1$. Hence the $P$ value of a ferromagnetic metal can be determined using a known singlet SC.

For a triplet SC, AR spectrum of a normal current ($P = 0$) is the same as that of a singlet SC, regardless of the spin states in the triplet SC. The triplet SCs of most interest in quantum computing [21] and spintronics [22] are those that can carry spin angular momentum. For these triplet SCs, AR can occur even for a half-metallic current. As a result, conductance within the gap is not zero, but depends on the exact spin states and the gap symmetry [24]. Therefore, ARS with a highly spin-polarized current can determine the spin states of the Cooper pairs. The above discussion is for an ideal interface. Experimentally, the interface is often not ideal and there are interfacial scattering ($Z$), inelastic scattering ($\Gamma$), temperature ($T$), and extra resistance ($r_E$) [25]. These effects can be described by a modified Blonder-Tinkham-Klapwijk (BTK) model [25] for an *s*-wave SC. The gap ($\Delta$) and $P$ values are extracted by fitting the whole ARS spectrum to the modified BTK model. Spin polarization of many magnetic materials [26-38] including some highly spin-polarized [28-35] and half-metallic materials [28-29] has been determined by ARS using a known singlet SC. Similarly, superconducting gap of many SCs, including two band $MgB_2$ [39], and iron SCs [38, 40] has been studied by ARS using metals with known $P$ value.

Coexistence of ferromagnetism and superconductivity has been previously observed in NiBi$_3$ nanostructures [17]. But later experiments did not find ferromagnetism in NiBi$_3$ single crystals [18-20]. It has also been shown that Bi/Ni bilayers have *p*-wave triplet superconductivity [42, 43]. And NiBi$_3$ alloy can be formed between Bi/Ni bilayers growing at 60 °C [44]. Therefore, it is important to determine the nature of superconductivity in NiBi$_3$ alloy in addition to ferromagnetism. In this work, ARS with an unpolarized current has been utilized to determine the gap value, structure and *T* dependence of the intermetallic single-crystal NiBi$_3$. A highly spin-polarized current injected from La$_{2/3}$Sr$_{1/3}$MnO$_3$, (LSMO) has been used to determine spin states. The spin states of the polycrystalline NiBi$_3$ sample is antiparallel, the same as that of a singlet SC. The critical field of the sample is about 150 Oe, observed from ARS in a small magnetic field. The point contact shows a negative magnetoresistance (MR) for unpolarized current but a positive MR for a spin-polarized current, demonstrating the distinctive difference of AR for unpolarized and polarized current of a singlet SC. The gap value $2\Delta/k_B T$, symmetry and its *T* dependence, combined with the spin states show that the single-crystal NiBi$_3$ is a singlet s-wave superconductor.

**Experimental details**

The NiBi$_3$ single crystals were grown from self-flux method [19]. In brief, high purity Ni shot and Bi pieces were mixed and sealed in an evacuated quartz tube with molar ratio Ni:Bi = 1:10. The tube was heated up to 1150 °C and soaked for 2 h, then cooled down to 400 °C with a cooling speed of 5 °C/h. Finally, the tube was spun in a centrifuge to separate the extra Bi flux. This method avoided not only the NiBi phase, also possible magnetic Ni impurities, and successfully realized the stoichiometric NiBi$_3$ single crystal. The detailed growth of the sample along with the characterization of the single crystal NiBi$_3$ using XRD, transport and Meissner effect has been described previously [19] where lattice parameters as well as crystal directions have been determined for the sample. The *Tc* of the single-crystal sample is about 4.07K as shown in Fig. 1 from the temperature dependence of resistance. The single crystal sample La$_{2/3}$Sr$_{1/3}$MnO$_3$ (LSMO) was grown by floating-zone method described before [45].

The needle-like NiBi$_3$ single crystal samples are of dimensions of ~3 mm× 0.2 mm× 0.2 mm with *b* axis along the needle. For ARS experiment, the sample has been cut or bent to contact on different crystalline directions. The sample is enclosed into a vacuum jacket along with a polished Au surface with 99.999% purity or a single crystal LSMO surface. The sample tube is pumped to 10$^{-7}$ Torr then filled with 0.1 Torr of Helium gas. The samples are then cooled down to 4.2 K using liquid helium. Lower temperature is achieved by pumping the sample tube. At desired temperature, a point contact is established by approaching the tip to the surface via a differential screw mechanism. The differential conductance *dI/dV* and resistance *V/I* are measured simultaneously using a lock-in method. For a new contact, the NiBi$_3$ tip is cut or bent again to obtain a new tip.

**Results and Discussions**

First a point contact is established on nonmagnetic Au using the NiBi$_3$ tip. A current from Au is unpolarized, the conductance within the gap Δ should be doubled for an ideal interface, regardless of the singlet or triplet nature of the NiBi$_3$ sample. But a real interface often has interfacial scattering Z which causes a dip at *V* = 0. Some representative Andreev spectra are shown in Fig. 2(a-e).  The open circles are the experimental data and the solid curves are the best fit to the modified BTK model.  In our fitting, only Δ, Γ and Z are varied while $T$ is set as experimental value and $P$ is 0.  The inelastic scattering factor Γ is close to zero (<0.001) for all contacts indicating the high quality of the NiBi$_3$ sample.  The data can be well described by the model and the parameters of the best fit are listed as inset.  The conductance rises up to 1.9, very close to an ideal interface with a small Z factor of 0.15, as shown in Fig. 2(a). For increasing Z factor, the conductance decreases and there is a dip appears at zero-bias voltage for large Z factor, as shown in Fig. 2(e).  The ARS spectra are very different for different Z factor, nevertheless the determined Δ values are very similar.  We have measured over 10 spectra using several NiBi$_3$ tips and all the determined gap values are similar.  The average Δ value is 0.62 ± 0.01 meV, as shown in Fig. 2(f).  The ratio, $2\Delta/k_B T_c$ = 3.51, is very close to the BCS *s*-wave ratio of 3.53.  Since each contact from different tip may reach a different crystalline surface of the NiBi$_3$ sample, the similar gap value of all contacts indicate an isotropic gap for the NiBi$_3$ sample.

  After the determination of the gap value and its symmetry with an unpolarized current, next we utilize a highly spin-polarized current to determine the spin states of the Cooper pairs in NiBi$_3$.  The polarized current is injected from LSMO, which has a spin polarization over 80% determined by singlet SC Pb [38].  Some representative ARS spectra from NiBi$_3$/LSMO contact are shown in Fig. 3(a-e).  The open circles are the experimental data and the solid curves are the best fit to the modified BTK model.  All the spectra can be well-described by the modified BTK model and the parameters of the best fit are listed as inset.  In analysis, the gap is fixed as 0.62 meV determined above and $T$ is set as experimental values. Only $P$, Z and $r_E$ are varied. The extra resistance $r_E$ is utilized to incorporate the effect of the large resistivity from LSMO. The effect of $r_E$ in ARS has been previous studied in details [37]. The spectra are very different from that of $NiBi_3$/Au contacts in Fig. 2: The conductance at zero bias is much lower than 1 due to the suppression of AR by spin polarization. Indeed, the obtained $P$ value is about 80% at small Z factor, as shown in Fig. 3(a). For increasing Z factor, the shoulder peaks increase, as shown in Fig. 3(e). The $P$ values obtained from different contacts are plotted in Fig. 3(f). For increasing Z, the $P$ value decreases due to spin-flipping scattering at the interface. The intrinsic spin polarization of LSMO, $P = 0.812 \pm 0.019$, is obtained by extrapolating Z to 0. If there is any spin-parallel triplet pairing with $S_z = 1$ in the $NiBi_3$ SC, the $P$ value will be zero because AR of a triplet SC is not suppressed by a spin-polarized current. The $P = 0.81$ is the same as that obtained using a singlet SC [38], suggesting that all spin states in the $NiBi_3$ sample is antiparallel, the same as that of a singlet SC.

The critical field of $NiBi_3$ is small, about 150 Oe and this can also be observed in ARS. As shown in Fig. 4(a), the ARS spectrum of a $NiBi_3$/Au point contact can be well-described by the modified BTK model. In this analysis, only $T$ is fixed as experimental value but $P$, $\Delta$, and Z are varied. The obtained $P$ is zero for gold. Then we apply a magnetic field ($H$) and measure the magnetoresistance (MR) at small bias of 0.1 mV, which is less than the gap value. As shown in Fig. 4(b), it shows a negative MR. At H < 150 Oe, the resistance is smaller, and it jumps to a higher value when the $NiBi_3$ sample becomes normal. When the $NiBi_3$ sample is superconducting, AR occurs thus conductance is doubled or resistance is reduced by half. One notes that the resistance of the whole $NiBi_3$ sample at normal state is only about 0.1 $\Omega$ at room temperature. So the change of the resistance of about 1 $\Omega$ is mainly due to the disappearance of AR at H > 150 Oe.

For spin-polarized current, however, the MR is completely different. As shown in Fig. 4(c), the ARS spectrum of a NiBi$_3$/LSMO contact can be well described by the modified BTK model when all the parameter except $T$ are varied in the analysis. The obtained $\Delta$ and $P$ values are consistent with the value in Fig. 2 and 3. The MR at bias $V$ = 0.1 mV is positive, completely different from that observed using the Au contact in Fig. 4(b). The change of the resistance at $H$ = 150 Oe is about 80 $\Omega$, much large than the sample resistance of 0.1 $\Omega$. This MR is due to the suppression of AR from the highly spin-polarized current. If it is a true half metal of 100%, the resistance at zero bias would be infinite because of the suppression of AR. In the case of LSMO, the $P$ value is about 80%, but still, the change of resistance is very significant. These results further demonstrate that the spin states in the NiBi$_3$ sample must be antiparallel.

The MR described above is due to Andreev reflection, very different from MR in magnetic nanostructures such as giant magnetoresistance (GMR) [46, 47] or tunneling magnetoresistance (TMR) [48]. Sharp change of resistance (MR) causing by a magnetic field can be utilized as field sensors, e.g. read head in magnetic recording [49] and biosensors [50]. The sensitivity of a MR sensor depends crucially on two factors, the sharpness of the resistance jump and the MR ratio. In GMR and TMR, the sharp change occurs at the switching of the magnetization and the ratio depends on the spin-dependent scattering or tunneling process. In the Andreev MR, the sharp change occurs at the critical field of the superconductor and the MR ratio depends on the spin polarization and the contact resistance. The Andreev MR provides a new type of magnetic field sensor where the sensitivity can be tuned or improved very differently. Ideally, a half metal can induce an infinite ratio in TMR, but a real TMR junction depends crucially on the tunneling barrier of 1-2 nm. Half metal $CrO_2$ has been shown to have $P$ = 98% [28, 29], with an estimated Andreev MR of 2500% from conductance, but high TMR has not been observed in $CrO_2$ based magnetic tunneling junctions.

Although Andreev MR must work at low temperature, it has another advantage that is absent in GMR or TMR sensors. It can also sense electric field. As shown in Fig. 4(c), at the superconducting gap, the conductance change significantly at the gap of 1.3 mV. This big change depends again crucially on the $P$ value of the metal. So the sensitivity for both electric field and magnetic field can be improved by using highly spin-polarized materials.

Next, we measure the $T$ dependence of the gap. As shown in Fig. 5(a), one AR spectrum is measured from 1.47 K to 4.5K with about 0.2 K of increase for each curve. For increasing $T$, the AR spectrum decreases and the peak disappears exactly at the $T_c$ of $NiBi_3$ at 4.1K. One notes that the Hallmark double AR peaks disappear first, and the AR spectrum becomes a big single peak, then the single peak starts to decrease, as clearly shown by the 2D graph in Fig. 5(b), which is the same data of the 3D graph in Fig. 5(a). We analyze the AR spectrum at each $T$. For the spectrum at 1.47 K, we varied $\Delta$ and Z, and set $T$ = 1.47 and $P$ = 0. For higher $T$, since the Z factor of the same contact is not expected to change during the total change of $T$ less than 3 K, we fix the Z factor as 0.21, only vary the $\Delta$ value in analysis of each $T$. The obtained $\Delta$ values at different $T$ is plotted in Fig. 6(b), along with a BCS theory (dashed curve). One can see that most of the data are very close to the BCS model except a few points near the transition temperature, which may be due to the pressure on the point contact.

$NiBi_3$ SCs have been studied before. While it has been found that ferromagnetism and superconductivity can coexist in $NiBi_3$ nanostructures [17], ferromagnetism has not been found in bulk single crystals [18-20]. In this work, the gap value and symmetry, spin states and $T$ dependence of the gap all demonstrate that the bulk NiBi$_3$ sample is a singlet *s*-wave SC. However, this cannot rule out the possibility that triplet superconductivity exists in NiBi$_3$ nanostructures. Indeed, studies have shown that while ferromagnetism is absent in the bulk NiBi$_3$ single crystal there is ferromagnetic fluctuations at the Ni/Bi surface [19]. In the Bi/Ni bilayers or nanostructures, the effect of interface/surface is maximized, which can lead to triplet superconductivity. Tunneling experiment has shown that the current is spin polarized [26]. Nevertheless, whenever ferromagnetism or triplet superconductivity occurs, the critical field of the structure is several order of magnitude larger the critical field of bulk NiBi$_3$. Recently it has shown that the superconductivity in 2D materials depends crucially on interface or substrate [51]. While a bulk ferromagnetic SC is elusive, the results suggest that the future triplet SC may be engineered using interfaces in hybrid structures, where ferromagnetism can impact superconductivity.

In summary, Andreev reflection spectroscopy with unpolarized and highly spin-polarized currents has been utilized to study intermetallic single-crystal superconductor NiBi$_3$. The gap is isotropic with a value of 0.62 meV and $2\Delta/k_BT_c$ = 3.51. The spin state is antiparallel, the same as that of a singlet superconductor. The temperature dependence of the gap displays a BCS-like behavior. The gap value, isotropic symmetry, and its temperature dependence, along with the antiparallel spin state conclusively demonstrate that the bulk NiBi$_3$ crystal is an *s*-wave singlet superconductor. In addition, an Andreev magnetoresistance has been observed in point contacts with unpolarized and spin-polarized currents and it is negative for unpolarized currents but positive for highly spin-polarized currents.

Acknowledgments:

This work was supported as part of SHINES, an EFRC center funded by the U. S. Department of Energy, Office of Science, Basic Energy Science, under award SC0012670.

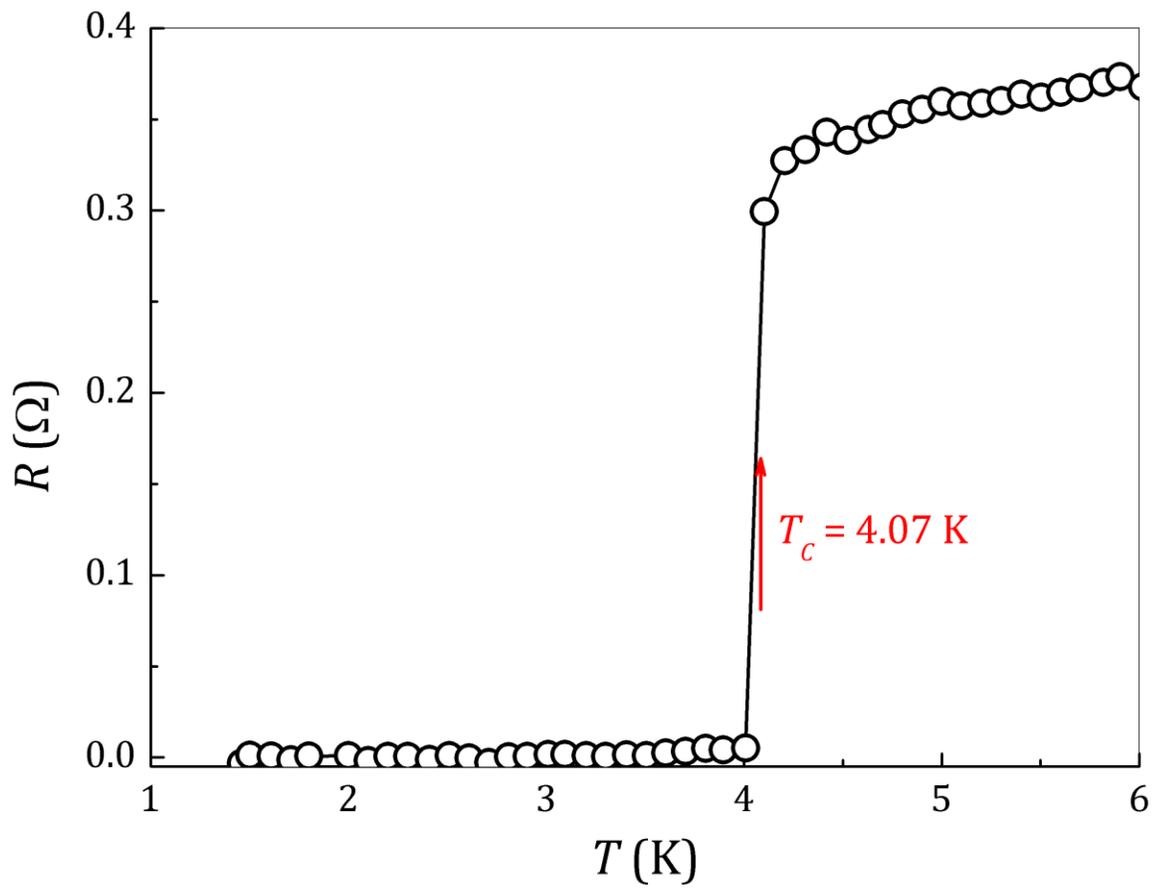

Fig. 1 Resistance of a NiBi$_3$ single-crystal sample as a function of temperature.

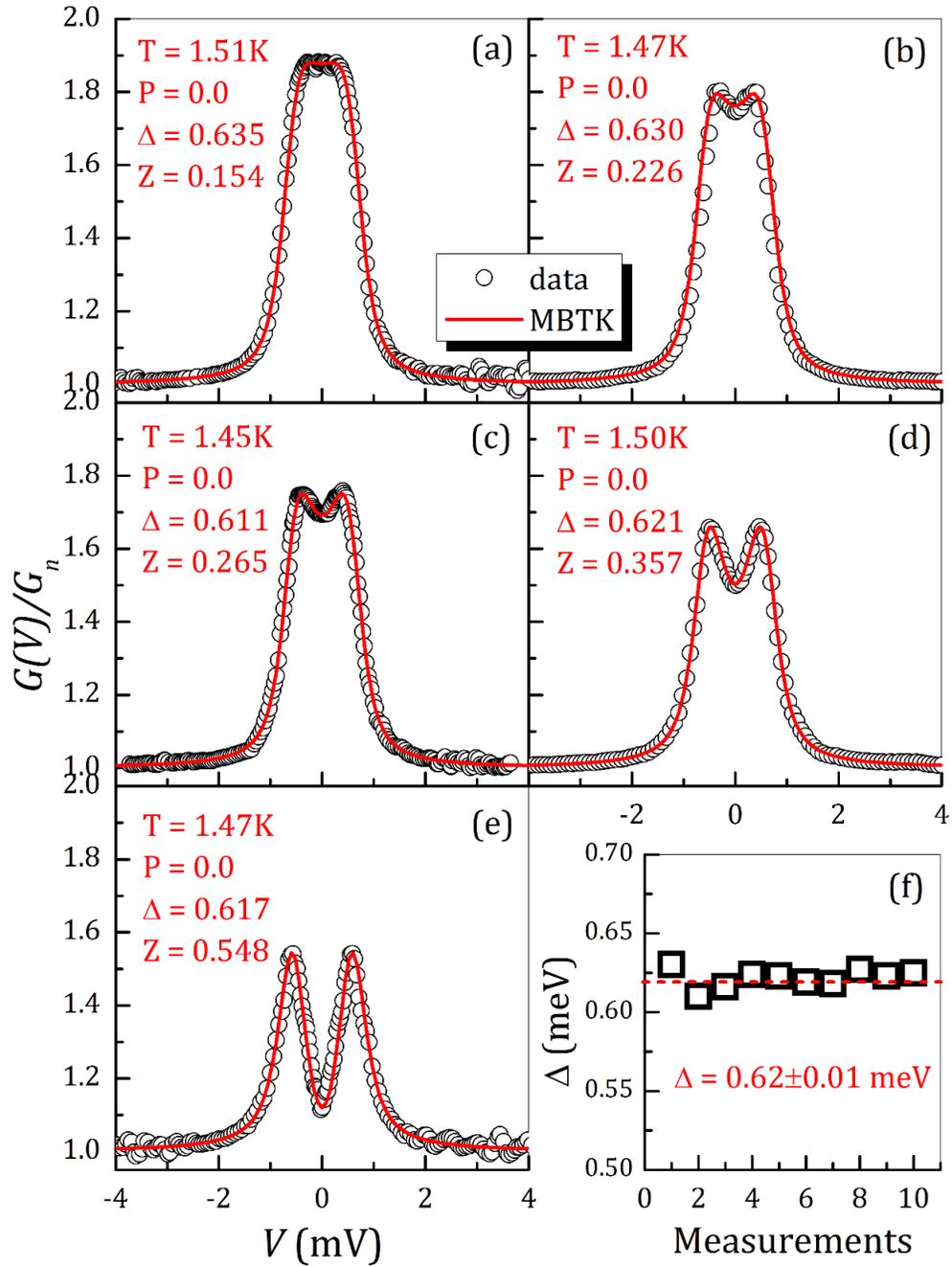

Fig. 2(a-e) Representative Andreev spectra of NiBi$_3$/Au contacts with various interfacial scattering Z factor (Open circles are the experimental data and the solid curves are the best fit to the modified BTK model), and (f) gap values determined from different contacts.

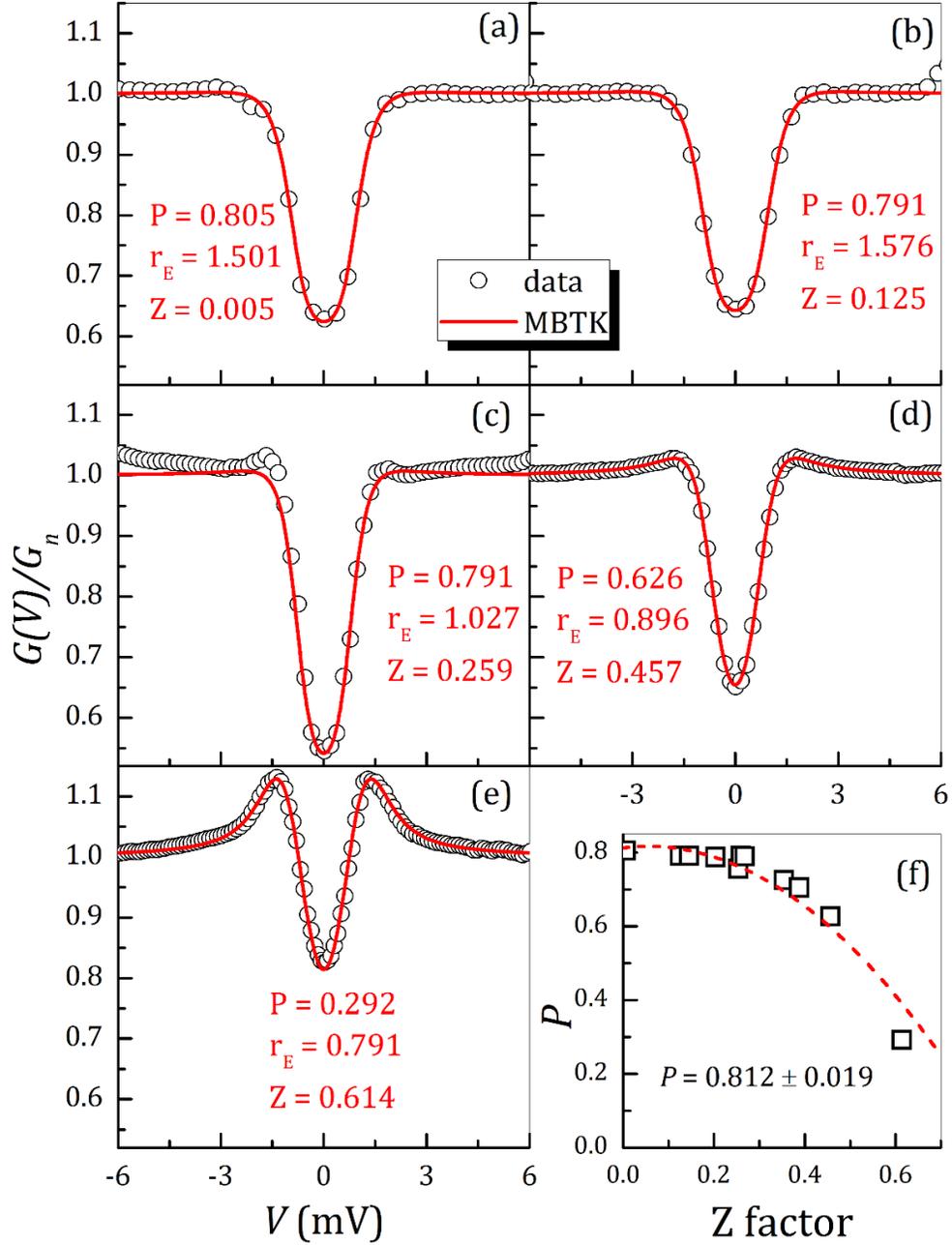

Fig. 3(a-e) Representative Andreev spectra of NiBi$_3$/LSMO contacts with various interfacial scattering Z factor (Open circles are the experimental data and the sloid curves are the best fit to the modified BTK model), and (f) spin polarization *P* as a function of interfacial scattering factor Z.

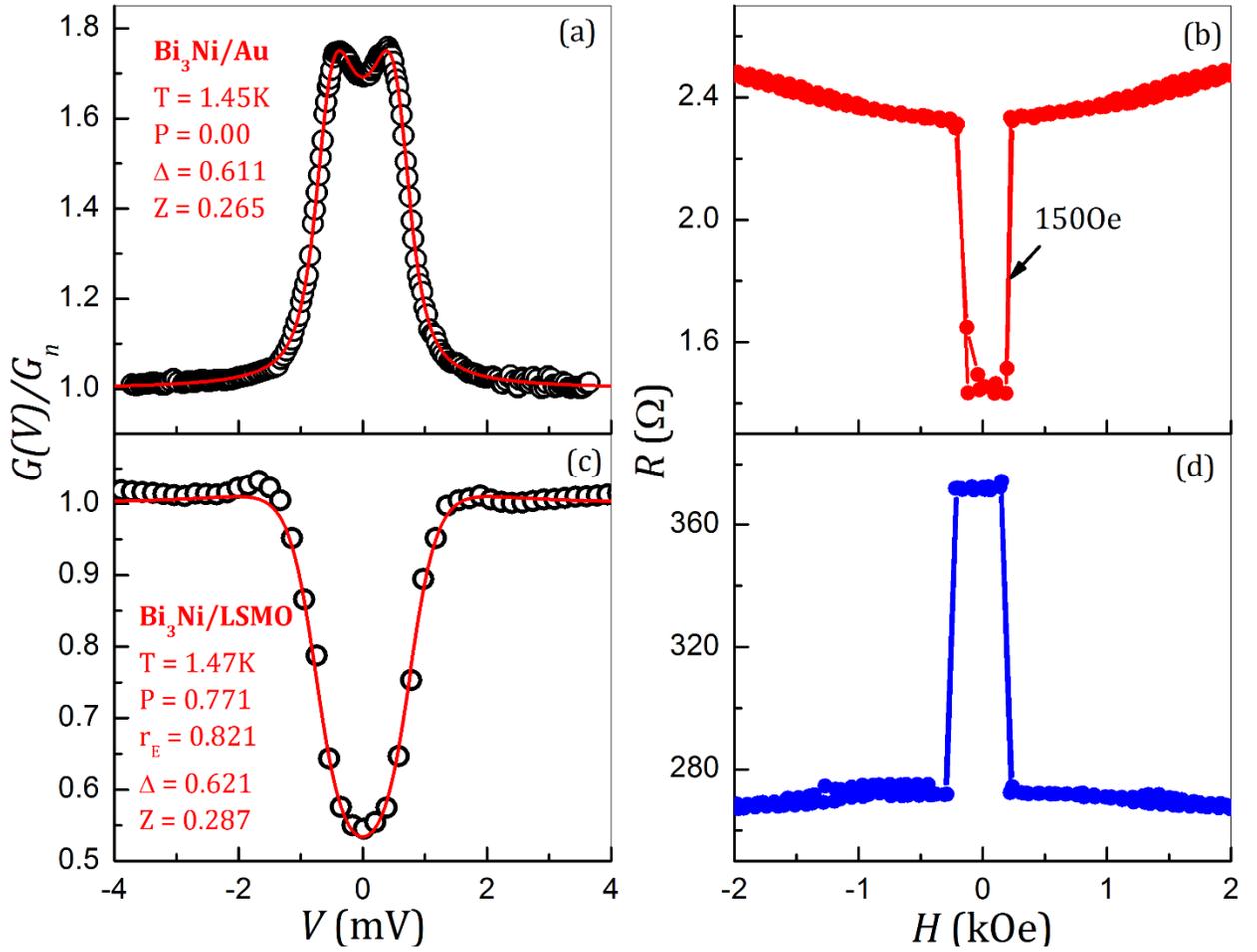

Fig. 4(a) Andreev spectrum of one NiBi$_3$/Au contact and (b) magnetoresistance at $V \approx 0$ of the contact, (c) Andreev spectrum of one NiBi$_3$/LSMO contact and (d) magnetoresistance at $V \approx 0$ of the contact.

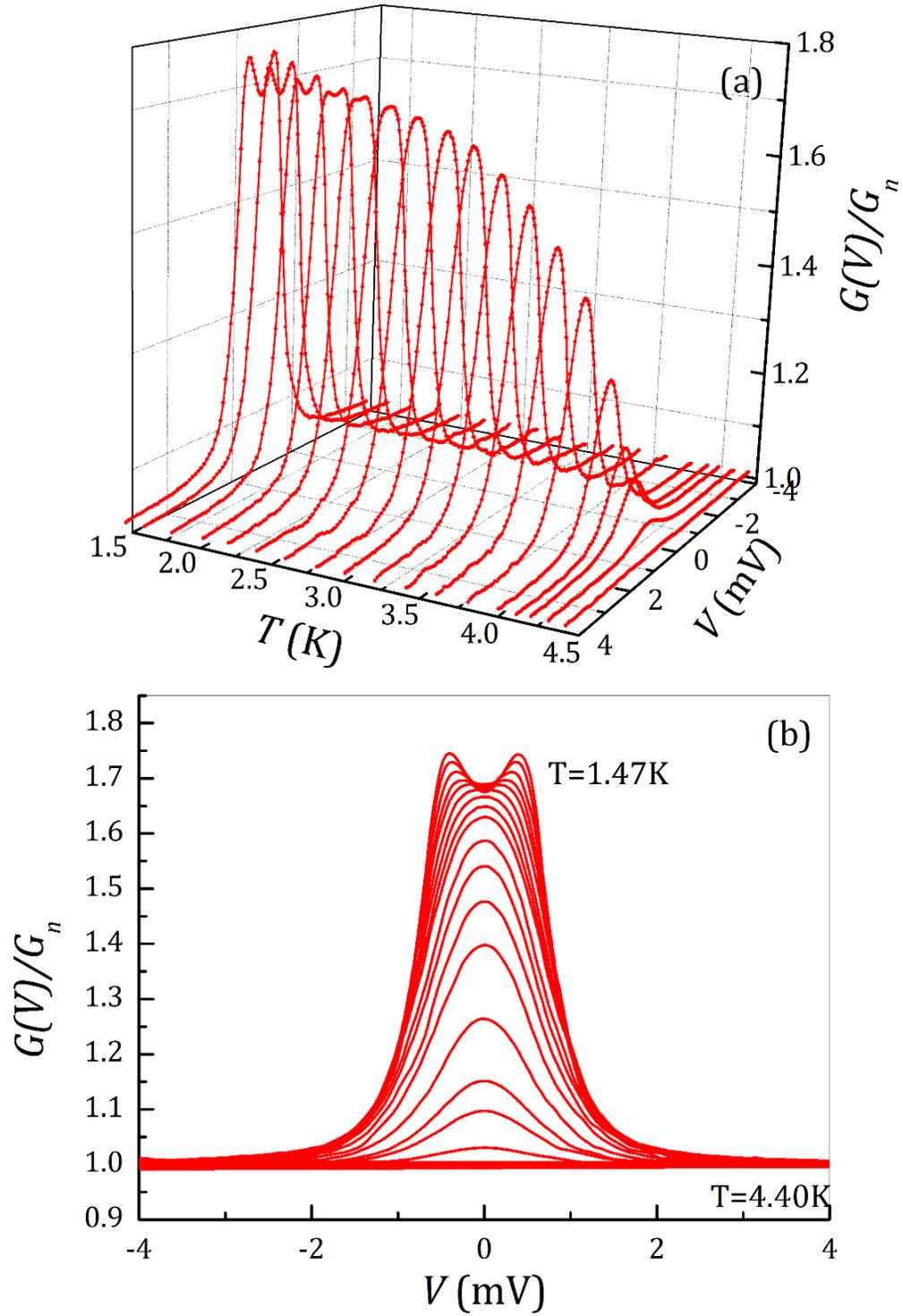

Fig. 5 Temperature dependence of one Andreev spectrum of a NiBi$_3$/Au contact from 1.47 K to 4.5 K with a temperature step about 0.2 K in (a) 3D plot and (b) 2D plot.

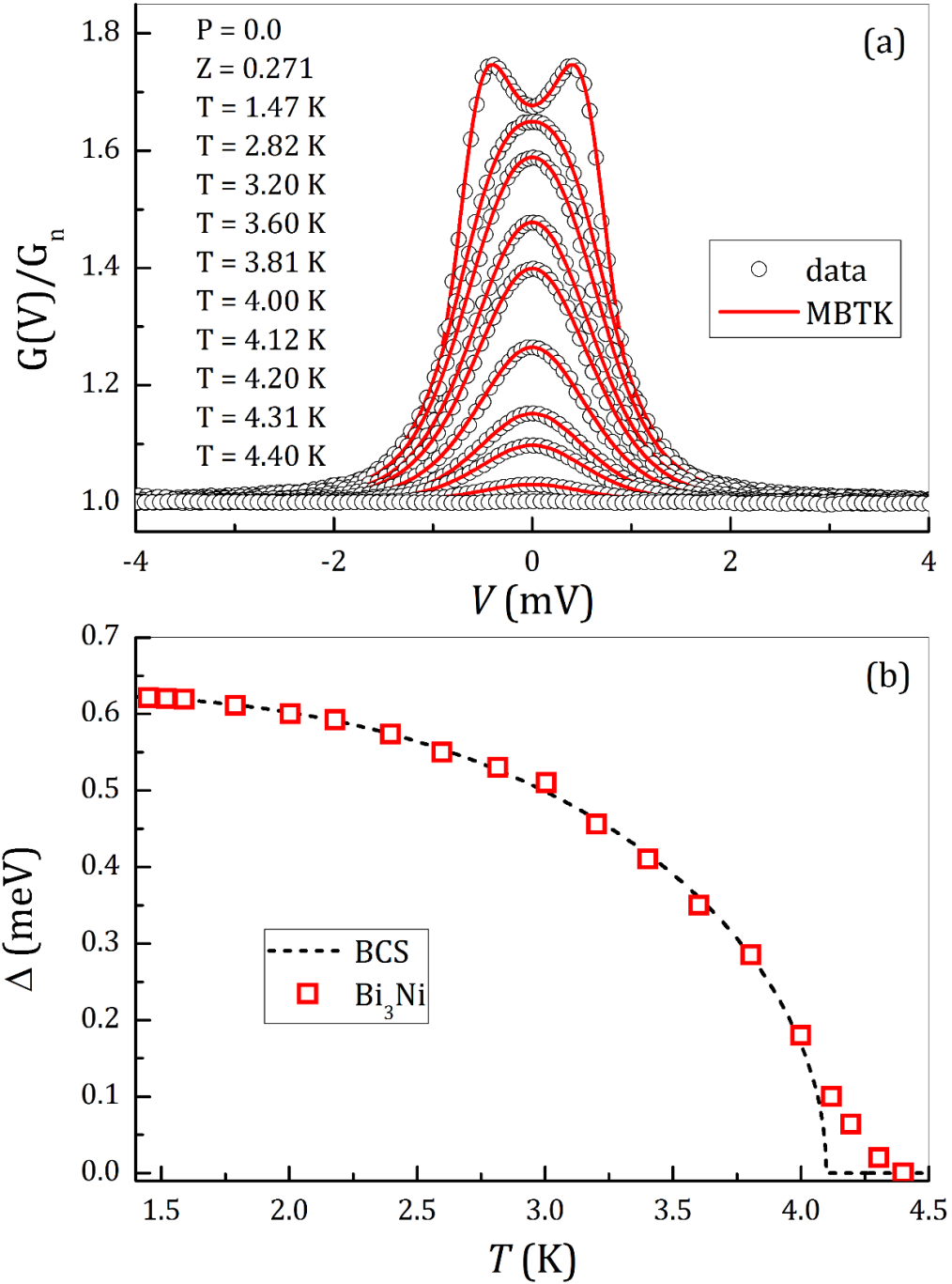

Fig. 6(a) Representative Andreev spectra (open circles) and the best fit (solid curves) at different temperatures, and (b) gap values obtained from the best fit at different temperatures from 1.47 K to 4.5 K with Z = 0.271 (open square), and the BCS gap dependence (dashed curve).